# Continuous Double Auction Mechanism and Bidding Strategies in Cloud Computing Markets

Xuelin Shi, Ke Xu, *Senior Member, IEEE*, JiangChuan Liu, *Senior Member, IEEE*, Yong Wang

**Abstract**—Cloud computing has been an emerging model which aims at allowing customers to utilize computing resources hosted by Cloud Service Providers (CSPs). More and more consumers rely on CSPs to supply computing and storage service on the one hand, and CSPs try to attract consumers on favorable terms on the other. In such competitive cloud computing markets, pricing policies are critical to market efficiency. While CSPs often publish their prices and charge users according to the amount of resources they consume, auction mechanism is rarely applied. In fact a feasible auction mechanism is the most effective method for allocation of resources, especially double auction is more efficient and flexible for it enables buyers and sellers to enter bids and offers simultaneously. In this paper we bring up an electronic auction platform for cloud, and a cloud Continuous Double Auction (CDA) mechanism is formulated to match orders and facilitate trading based on the platform. Some evaluating criteria are defined to analyze the efficiency of markets and strategies. Furthermore, the selection of bidding strategies for the auction plays a very important role for each player to maximize its own profit, so we developed a novel bidding strategy for cloud CDA, BH-strategy, which is a two-stage game bidding strategy. At last we designed three simulation scenarios to compare the performance of our strategy with other dominating bidding strategies and proved that BH-strategy has better performance on surpluses, successful transactions and market efficiency. In addition, we discussed that our cloud CDA mechanism is feasible for cloud computing resource allocation.

## 1 Introduction

With the development of Information and Communication Technology, computing will one day become the 5$^{th}$ utility (after water, electricity, gas, telephony) [1]. Computing resources are always distributed dispersedly, which are connected with networks. How to provide transparent computing services for users in such a heterogeneous environment is a key problem. To deal with it, a number of computing paradigms have been proposed: cluster computing, Grid Computing, and more recently cloud computing. Cloud computing has been an emerging model which aims at allowing customers to utilize computational resources and software hosted by Cloud Service Providers (CSPs) [2].

There are many famous CSPs, such as Amazon, Google, Microsoft, Rackspace, Joyent, GoGrid, OpSource, Verizon/Terremark, Citrix, Bluelock, and so on. More and more consumers rely on CSPs to supply computing service on the one hand, and CSPs try to attract consumers on favorable terms on the other. Take Amazon for example, it has been estimated that Amazon EC2 has more than 70,000 EC2 server instances per day in 2010 [3]. At the same time, the scale of CSPs is also increasing greatly.

Obviously the cloud computing utility services have created a competitive open market environment. Every market participant searches for its own path to the maximum profit, while a market pricing mechanism should be applied to balance supply and demand in real time and maintain the market reliability. Just like markets for network resources, the pricing problem has become increasingly urgent. It is clear that if the markets are not properly designed, they could function rather poorly, even leading to market failure [4]. The traditional Internet was just a best-effort service without economic resources allocation, which resulted in poor network utility and congestion. Therefore many economic and technical approaches for network resources scheduling are brought, such as network utility maximization (NUM), network resources auction, time-dependence pricing, etc.

As cloud computing is designed to be a market-oriented computing paradigm, the effective market model and pricing mechanism both are critical for cloud computing to avoid previous Internet problems.

Now CSPs have their own pricing policies. Usually CSPs specify their service price and charge users according to the amount of resources they consume, which can be called posted-offer pricing model. The pricing policy

- *Xuelin Shi is with the Department of Computer Science and Technology, Tsinghua University, Beijing, China. E-mail: shixuelin@sina.com.*
- *Ke Xu is with the Department of Computer Science and Technology, Tsinghua University, Beijing, China. E-mail: xuke@mails.tsinghua.edu.cn.*
- *Jiangchuan Liu is with the School of Computing Science, Simon Fraser University, British Columbia, Canada. E-mail: jcliu@cs.sfu.ca.*
- *Yong Wang is with the School of Social Sciences, Tsinghua University, Beijing, China. E-mail: wang.yong@tsinghua.edu.cn.*

can be derived from various parameters. For example, Amazon EC2 provides different purchasing options, such as on-demand instances (paying for computing capacity by the hour with no long-term commitments, *$0.060 per Hour* for a small/default instance), reserved instances (making a low, one-time payment for each instance you want to reserve and in turn receiving a significant discount, *$0.034 per Hour* for a small/default instance), etc. Such a posted-offer pricing model can not incent users to consume more resources in off-peak periods or pay more in peak periods. Bidding and auction are effective ways to solve the above problem. However the auction mechanism is rarely adopted by CSPs. Only Amazon EC2 provides spot instances allowing customers to bid on unused Amazon EC2 capacity and run those instances as long as their bid exceeds the current spot price. It is one single-sided auction model, but more efficient auction models can be applied to cloud markets.

Furthermore, owing to the Internet, the rapid development of e-commerce has effectuated an increase in the number of users engaged in electronic auction (e-auction) services. E-auctions enable bidders to bid for diverse objects on an electronic platform via Internet [5]. For the users often access cloud services via Internet, E-auction is also a feasible method to solve cloud resources pricing problem.

Consequently, an e-auction platform can be established, on which many CSPs and users trade computing and storage resources online. Moreover the double auction mechanism is more competent for cloud markets than single-sided auction mechanisms adopted by CSPs at present. The buyers and sellers can both submit bids in double auctions. In the double auctions, the selection of bidding strategies plays a very important role for each player to maximize its own profit.

This paper makes two major contributions. First, a cloud Continuous Double Auction (CDA) mechanism is formulated to match orders and facilitate trading based on an electronic bidding platform. Second, we develop a novel bidding strategy for cloud CDA, BH-strategy, which is a two-stage game bidding strategy. At last we evaluate the efficiency of our strategy.

The following of this paper is structured as follows. In section 2 the research works on Internet Pricing, Continuous Double Auction (CDA) and bidding strategies are reviewed. Then we discuss challenges to cloud markets and bring up the conclusion that the e-Business platform is a feasible solution to cloud markets. We design a cloud CDA mechanism and e-bidding platform scheme in section 4. Section 5 describes our novel bidding strategy for cloud CDA. The simulation results are given in section 6. Section 7 concludes the paper.

## 2 Related Works

How to manage and schedule resources economically is very important in cloud computing. It has also been researched in grid computing, and some economy resource allocation frameworks have been proposed [6]. These works can be regarded as an application and extension of some market economy theories in computing resources markets, the key point of which is to set prices of the commodities in the markets.

Now there are less research works on cloud computing pricing than on Internet pricing. But some solutions to the latter can be applied to cloud computing, because the allocation of cloud computing resources is similar to that of network resources. Many network pricing policies are based on auction mechanism, which are proved efficient. In this section we will review Internet resource pricing first, and then present related works on auctions.

### 2.1 Internet Resource Pricing

The basic problem in communication networking is how to effectively share resources. Dynamic decentralized algorithms for network flow optimization have a long history, the mathematical theory under which is often referred to as network utility maximization (NUM). But it has arbitrarily bad efficiency when users behave selfishly and strategically. Thus, there is a need to go beyond the current distributed optimization-based NUM framework to a game-theoretic and market economics based network market design (NMD) framework that takes incentive issues into account in the design of network resource allocation algorithms and protocols.

Recently, with a tremendous growth in demand for broadband data, ISPs are forced to use pricing as a congestion management tool. There are flat-rate pricing, usage-based pricing, application-based pricing, time-dependent pricing and so on. But pricing policies with the right incentives to shift a demand from congested periods to off-peak times are more popular. Such peak-load pricing, off-peak discounts, day-ahead pricing, and real-time pricing have an explicit time-dependent pricing feature; others, like game-theoretic (auction) models and prioritization-based models (e.g., smart market, raffle-based, token bucket pricing), use more implicit time-varying incentives to induce the time-shifting of data to less congested periods [7].

Auction and market design solutions to network resources exist in a wide variety of scenarios. Single-sided auction designs for divisible goods have been explored fairly well. Designing suitable double-sided auctions, or markets, has proved to be rather challenging [4].

However, there are also differences between the Internet and cloud pricing. While the Internet is widely used by many common people, the cloud computing services are mainly for scientific computing and websites [8, 9]. Compared with common users, cloud users are usually more professional and willing to accept e-auction services. Therefore, the auction mechanism is feasible for cloud markets pricing.

### 2.2 Reviews of CDA

In a real world market, there are various economic mod-els for setting the price of services based on supply-and-demand and their value to users, including commodity market, posted prices, tender and auction models. Many research works have attempted to apply these economic models to grid computing. In the above models bidding and auction have high potential for computing resource allocation in gird or cloud environments. But there are few researches engaged to analyze CSPs bidding strategies on game theory approach, which is important problem in biding and auction.

The auction model supports one-to-many or many-to-many negotiations, between a service provider (sellers) and many consumers (buyers), and reduces negotiations to a single value (i.e. price). Auctions can be conducted as open or closed depending on whether they allow back-and-forth offers and counter offers. The consumer may update the bid and the provider may update the offered sale price. Depending on these parameters, auctions can be classified into the following types: English Auction (first-price open cry), first-price sealed-bid auction, second-price sealed-bid auction (Vickrey auction), Dutch auction and double auction.

Gode and Sunder [10] divided double auction into three categories: Synchronized Double Auction, Continuous Double Auction (CDA) and Semi-continuous Double Auction (or Hybrid Double Auction).

CDA is one of the most common forms of market-places and has emerged as the dominant financial institution for trading securities and financial instruments. Indeed, today the major exchanges, like the NASDAQ and the New York Stock Exchange (NYSE) and the major foreign exchange (FX), apply variants of the CDA institution [11]. Other significant applications are in market-based control, where CDAs provide a dynamic and efficient approach to the decentralized allocation of scarce resources [12].

In the CDA model, buy orders (bids) and sell orders (offers or asks) may be submitted at anytime during the trading period. If at any time there are open asks and bids that match or are compatible in terms of prices and requirements (e.g., quantity of goods or shares), a trade is executed immediately. In this auction, orders are ranked from the highest to the lowest to generate demand and supply profiles. From the profiles, the maximum quantity exchanged can be determined by matching asks (starting with the lowest price and moving up) with demand bids (starting with the highest price and moving down). Researchers have developed software-based agent mechanisms to automate double auction for stock trading with or without human interaction [13].

Bidding strategy is an important problem in CDA, and many strategies for CDA have been subsequently developed. Over the last decade, there has been a considerable emphasis on strategies for software trading agents with the emergence of electronic markets [14], such as Zero-Intelligence (ZI) strategy [15], Zero-Intelligence Plus (ZIP) strategy [16], GD strategy [17], its subsequent extension GDX [18], p-strategy [19], k-Zero-Intelligence strategy [20], AA (Adaptive Aggressive) strategy [11], and so on.

Furthermore, with the development of Internet and e-Business, CDA is also adopted by many e-auction sites, such as FastParts, LabX, Dallas Gold and Silver Exchange.

If a CDA mechanism can be designed for cloud markets, it is also can be implemented on an e-auction platform. In cloud markets, the commodities can be purchased and delivered over the Internet [21]. Cloud users often have a variety of application and valuation types [22]. Rather than letting resources sit idle, CSPs are also inclined to sell unused resources at a reduced price via using auctions to users. Because of the above characteristics of cloud, online e-auction is an effective method to implement computing resources allocation. Especially, the double auction mechanism is more flexible for it supports buyers and sellers to bid simultaneously. However, double auction mechanisms have not been applied in cloud markets up till the present moment.

## 3 Challenges to Cloud Markets

Cloud Computing is emerging nowadays as a commercial infrastructure that eliminates the need for maintaining expensive computing hardware [23]. Different from Grid services which are billed using a fixed rate per service or different organizations sharing idle resources, cloud users are usually billed using a pay-per-use model [24]. Therefore market mechanism should be explored to facilitate trading between CSPs and users. In this section we analyze the challenges to the cloud markets and discuss why the CDA mechanism is a feasible method to resource allocation in such markets. At last we provide a scenario to describe how to apply the CDA in cloud markets.

Like other competitive markets, there are several challenges to cloud computing and storage resources market, including the following:

**1. Efficiency:** market efficiency can be evaluated by the ratio of the surpluses of all traders to the possible maximum surpluses that would be obtained in a centralized and optimum allocation. An efficient cloud market should obtain equilibrium of supply and demand while maximizing surpluses of buyers and sellers.

**2. Openness:** cloud is a market-oriented computing paradigm, which should allow everyone to consume computing and storage resources freely. Therefore a cloud market must be an open platform, which can be accessed by CSPs/users via the Internet.

**3. Fairness:** market rules must be fair to sellers and buyers, who can get rational surpluses from trades. A fair cloud market will regulate supply and demand of computing and storage resources, which can also facilitate the development of cloud computing. It not only concerns market rules, but also deals with tradeoffs of technical limits such as virtual machines, network, and so on [25].

**4. Feasibility:** an efficient and fair market mechanism designed for cloud markets should be feasible, which can be implemented as an e-commerce platform to help CPSs and users trading conveniently.

The above problems are also general requirements for many market mechanisms, including Internet service markets [4], wireless spectrum markets [26] and so on. Auction mechanisms are usually feasible to solve these problems, and the CDA is more efficient and fair than single auctions in cloud markets. Cloud users often have a variety of application and valuation types, while CSPs also have various idle resources. The double auction mechanism allows both users and CSPs to submit their demands/commodities. Furthermore, the CDA mechanism permits buyers and sellers bidding simultaneously in one auction. Advantages of the CDA for cloud markets can be revealed in the following scenario shown in Fig. 1.

For example, a user *Tom* has a computing job needing done in *10 hours* (i.e. the deadline of the job is *10 hours*) and is willing to pay *$5* for it (i.e. the budget of the job is *$5*). The job size is estimated to be *40 units*. One unit means resource capacity of one standard Virtual Machine instance running for one hour, which can be predefined as a regulation of the cloud electronic CDA platform. *Tom* can publish the demand of the job on the platform. The CSPs interested in the demand can submit asks no less than their unit costs for executing the job. *Tom* can not only wait for acceptable ask, but also submit his bids no more than *$5*. His first bid is maybe *$2*, and he submits the second bid *$4* if no CSPs accept his first bid. Similarly, if another user *Peter* has the same size and deadline job and his valuation for it is *$10*, he can submit his bids for it too.

During the auction, asks and bids are submitted continuously. Any new ask must be less than all current asks, and any new bid must be more than all current bids. When an ask is equal or less than a bid, a transaction occurs.

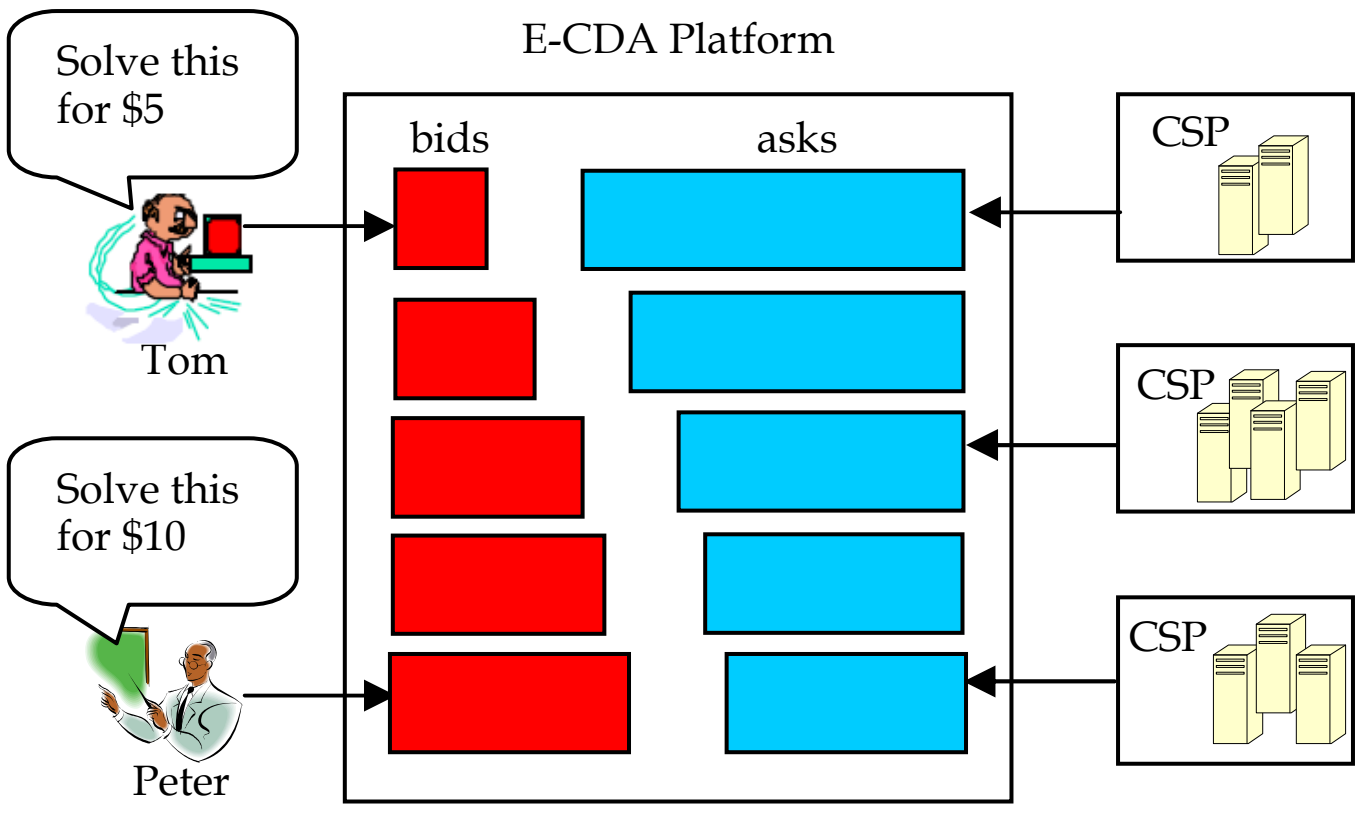

Fig. 1. Scenario of CDA in Cloud Computing Markets: Cloud Users and CSPs may submit bids and asks anytime.

In the other hand CSPs can also publish their idle resources on the platform. For example, a CSP *ACloud* has an idle storage space (*1TB*) for *6 months* (from *2013/6/1* to *2013/12/31*), and the cost is *$10* with PUT, COPY, POST, LIST, GET and all other requests less than *100,000*. *ACloud* can publish it on the platform. The users can bid for it, while other CSPs can submit asks of the same storage commodity.

Apparently the CDA is more efficient and flexible than single-sided auction in cloud markets, because it enables buyers to enter competitive bidders and sellers to enter competitive offers simultaneously. However, to apply CDA in real markets there are many detail rules and public parameters should be formulated. In next section a cloud CDA mechanism are presented.

## 4 A Cloud CDA Framework

The cloud computing market structure considered in this paper consists of CSPs, cloud users and the uniform bidding platform. Therefore a *complete competitive market* can be formed. This section presents the solution to resource allocation in such a market, including the model of cloud CDA mechanism and its market rules. Then an e-bidding platform scheme is proposed to implement the mechanism in real cloud environment. Our cloud CDA mechanism is an efficient way of the decentralized allocation for computing resources.

### 4.1 E-CDA Platform to Cloud Markets

For a huge cloud computing market populated by millions of users and CSPs, a uniform trading platform is vital. As stated above, an electronic bidding platform is a feasible solution, which can be easily accessed via the Internet and make use of e-Business technologies. Currently many kinds of commodities are traded on the e-Business platform, such as electric energy, petroleum, stock and so on.

On such an e-Business platform it is an efficient trading way that CSPs and users submit their orders simultaneously, so the CDA model can be applied. Such an e-bidding platform plays the role of an auctioneer, on which cloud users can submit buy orders (bids), while CSPs can submit sell orders (asks). To facilitate the trading the computing resources can be valued by a uniform unit, which is a homogeneous scenario.

Therefore we propose a cloud electronic auction platform, which apply the customized CDA mechanism to implement pricing and resource allocation in cloud markets. The essence of the e-CDA platform is how to formulate rules of the cloud CDA mechanism, as the following.

### 4.2 Model of Cloud CDA Mechanism

To formulate the cloud CDA mechanism, we firstly explore some of the basic notions:

**Definition 1.** *The* ***outstanding bid****,* $o_{bid}$*, is the current maximum demand order submitted by a cloud user in the market. The* ***outstanding ask****,* $o_{ask}$*, is the current minimum offer submitted by a CSP at any given time t in the market.*

**Definition 2.** *The **trading round**, **TR**, is the period during which asks and bids are submitted until there is a match and a transaction occurs. There are typically several trading rounds in a trading day. At the beginning of the trading round,* $o_{bid} = 0$ *and* $o_{ask} = Max$.

**Definition 3.** *The **bid-ask spread**,* $s_o$, *is the difference between* $o_{ask}$ *and* $o_{bid}$, $s_o = o_{ask} - o_{bid}$.

**Definition 4.** *The **limit price**, **LP**, is the maximum bid a cloud user is currently willing to pay, and the minimum ask a CSP is willing to supply.*

The cloud CDA mechanism can be thought as a discrete system: transforming a series of discrete input value (bids and asks) to a series of discrete output (transaction results, i.e. matching of bids and asks). Therefore the model can be described as the following:

$$M = F_{L_{TD},\Delta,Max,Min}(B,V,A,C) \quad (1)$$

**Public Parameters:**

$L_{TD}$: the **length of Trading Day**. A trading day, *TD*, is the period during which users and CSPs are allowed to submit offers and bids (resulting in transactions whenever they match), by the end of which the auction close.
$\Delta$: the **minimum increment** of a bid or ask in the market.
*Max*: the **maximum ask** allowed in the market, is to prevent unreasonably high asks and speed up the trading process.
*Min*: the **minimum bid** allowed in the market, is to prevent unreasonably low bids and speed up the trading process, usually set as 0.

**Input:**

$B = \{B_1, ..., B_i, ..., B_m\}$: **bid set** of *m* cloud users. $B_i$ is a subset containing all bids of cloud user *i*, and each bid is noted as $b_i^t$

$V = \{V_1, ..., V_i, ..., V_m\}$: **limit price set** of *m* cloud users. $V_i$ is the limit price of cloud user *i*, that is the highest bid it is willing to pay. Normally $V_j$ is user$_i$'s unit valuation for commodity, i.e. redemption value.

$A = \{A_1, ..., A_j, ..., A_n\}$: **ask set** of *n* CSPs. $A_j$ is subset containing all asks of CSP *j*, and each ask is noted as $a_j^t$

$C = \{C_1, ..., C_j, ..., C_n\}$: **limit price set** of *n* CSPs. $C_j$ is the limit price of CSP *j*, which is the lowest bid it is willing to submit. Normally $C_j$ is CSP$_j$'s unit cost for the production of commodity.

In this paper the cloud users are supposed to be rational. Therefore the bid of user *i* is no more than the price it is willing to pay, i.e.

$$Min \le b_i \le V_i \quad (2)$$

Accordingly the ask of CSP *j* is no less than the unit cost for the production of commodity, i.e.

$$C_j \le a_j \le Max \quad (3)$$

**Output:** *M*: the successful transaction matching result

In such a discrete-time system, a bid or ask is submitted at each step in the market. In this paper, the *trading day* is imposed as the auction closing after given steps.

At present we only study the cloud CDA mechanism and bidding strategy based on following two hypothesises:

**Hypothesis 1.** *The market is homogeneous, that hosts trading in a particular type of commodity where each unit traded is functionally identical to every other unit traded.*

**Hypothesis 2.** *The cloud users and CSPs have common knowledge of rationality and their orders must be subject to constraint (2) and (3), our cloud CDA is a model with constraints.*

To apply our cloud CDA mechanism in competitive cloud computing markets, we defined the market rules as followed:

**Rule 1.** *At each step, only one bid or one ask can be submitted. At any step t, if a bid or ask is submitted, then* $t=t+1$.

**Rule 2.** *Any new bid or ask must improve on the current outstanding bid or ask in the market, i.e.* $b^t > o_{bid}^{t-1}, a^t < o_{ask}^{t-1}$.

**Rule 3.** *At any step t, if* $o_{bid}^t \ge o_{ask}^t$, *then a transaction occurs at the price* $p_t = (o_{bid}^t + o_{ask}^t)/2$. *The winning cloud user's revenue is* $(V_i - p_t)$, *and the winning CSP's revenue is* $(p_t - C_j)$. *Then the winning buyer and seller are removed from the market. The current round is over, and next round begins.*

**Rule 4.** *At any step t, if the cloud user's (CSP's) limit price is lower (higher) than the current* $o_{bid}^t$ $(o_{ask}^t)$, *it cannot submit any bid (ask), and has to wait for the beginning of the next round. However, if it can submit a bid or ask in the cloud market, it considers its set of bidding strategies to form a price.*

**Rule 5.** *If* $t = L_{TD}$, *the trading is over.*

It is obvious that the bidding strategies play an important role in the trading. When a CSP decides a ask price, it must take the actions of other CSPs and all cloud users into account. And so does a cloud user. In section 4, we analyze the bidding strategies of cloud CDA.

### 4.3 Implementation of cloud CDA

The above cloud CDA mechanism is a universal model for resources allocation in cloud markets. In real cloud markets, the public parameters, $L_{TD}$, $\Delta$, *Max*, *Min*, can be regulated according to the requirements of markets.

Furthermore, current technology on cloud, especially Virtual Machines, makes it easy to divide computing re-sources to commodities. For example, a commodity in cloud markets maybe a time period of using a VM instance, some resource for running a scientific computing program in given time. Whatever a commodity is, CSPs can divide it to users.

The cloud CDA mechanism is practical in the real situation: cloud users (or user brokers) and CSPs can submit orders to an e-bidding platform via the Internet. We design such a scheme as shown in Fig. 3.

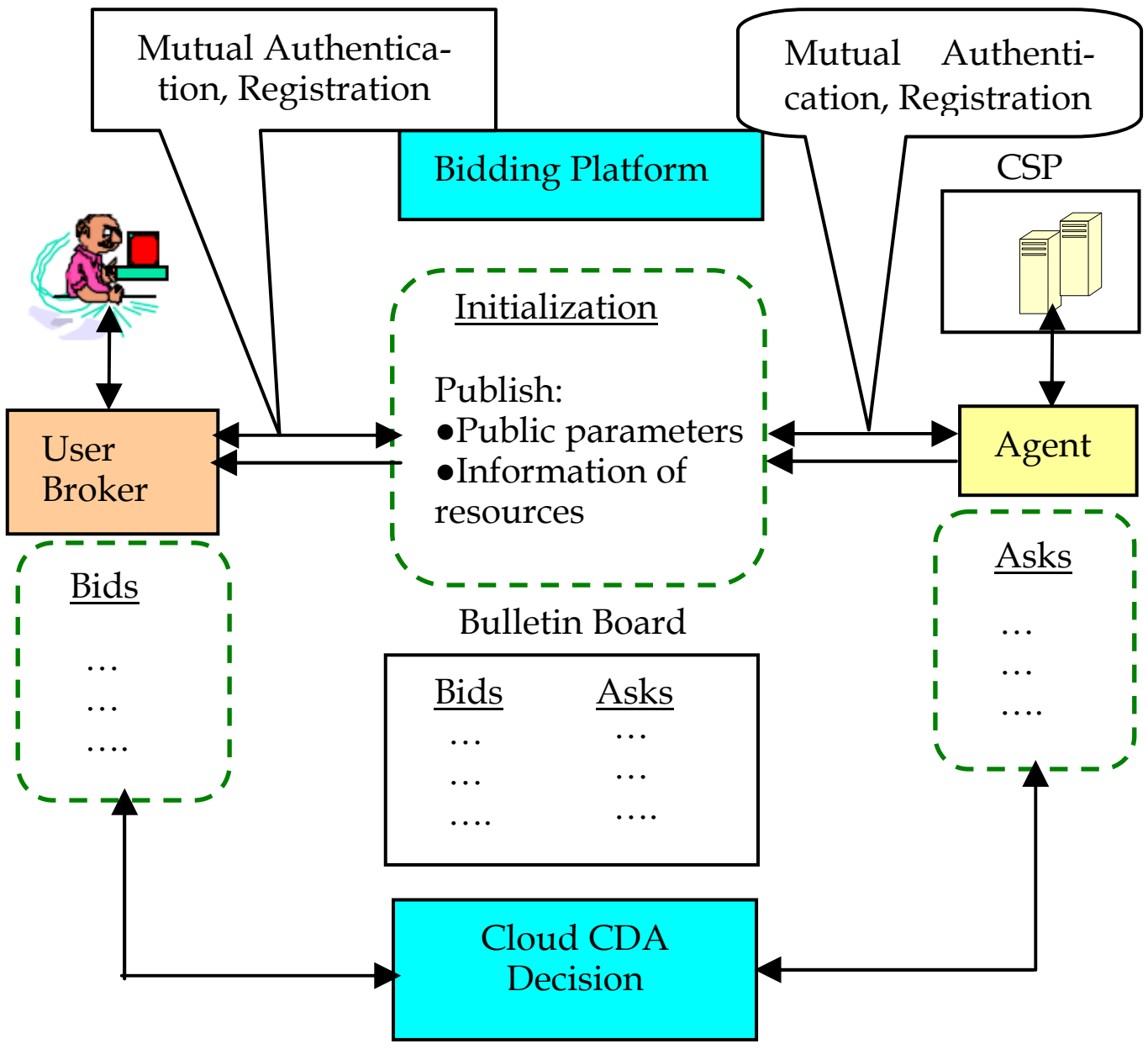

Fig. 3. Electronic Bidding Scheme in Cloud Computing Markets

The electronic bidding scheme in above figure is the public auction (outcry), where the bidding and asking prices of computing resources are public. For each bid, the lasted bid must be higher than all bids in the previous round of bidding. Similarly the most recent ask must be lower than all asks in the previous round. To facilitate cloud users' bidding, cloud users can authorize User Brokers (UBs) to bid for them on the e-bidding platform. CSPs usually have their agents in charge of bidding on the platform.

The auction on the e-bidding platform can be divided into 3 states: the registration stage, the bidding stage and the transaction decision stage. In the registration stage, the information of all computing resources and the related parameters of cloud users and CSPs are presented on the bulletin board, and every player is certified by the e-bidding platform. Then in the bidding stage, the UBs can submit bids and CSPs' agents can submit asks. Eventually, the highest bid and the lowest ask are matched by the cloud CDA decision module in the auction.

### 4.4 Evaluation Criteria

The objective of our cloud CDA mechanism is to maximize the profits of cloud users and CSPs. In such competitive markets, the market equilibrium occurs when demand meets supply in a free cloud computing market. According to classical micro-economic theory, the transaction prices in the CDA are then expected to converge towards that competitive equilibrium price $p^*$.

Therefore we design three criteria to evaluate the trade model and bid strategies. The market efficiency, $e_{market}$, is the ratio of the surpluses of all traders (i.e. cloud users and CSPs) to the possible maximum surpluses that would be obtained in a centralized and optimum allocation. The efficiency of a bidding strategy, $e_s$, is the ratio of the surpluses of traders adopting that strategy during a trading day to the maximum surpluses these traders could extract in a centralized allocation. In the homogeneous scenario, this is identical to the market efficiency, i.e. $e_{market} = e_s$. The daily price volatility, $\alpha$, shows how the transaction prices converge to the equilibrium price $p^*$. It is defined as:

$$\alpha = \frac{1}{p^*}\sqrt{\frac{\sum_{i=1}^{N}(p_i - p^*)^2}{N}} \quad (4)$$

As stated above, the cloud computing market is a competitive market, in which the competitive equilibrium price $p^*$ is decided by supply and demand together. The market-clearing is achieved at the equilibrium price $p^*$. Apparently $p^*$ cannot be known prior in this case because of the decentralized nature of the cloud CDA.

However to evaluate market efficiency (strategy efficiency) and price volatility, it is necessary to compute the approximation of the equilibrium price and the possible maximum surpluses, which would be obtained in a centralized and optimum allocation. This paper uses the Marshallian Path [27] to get an optimum allocation.

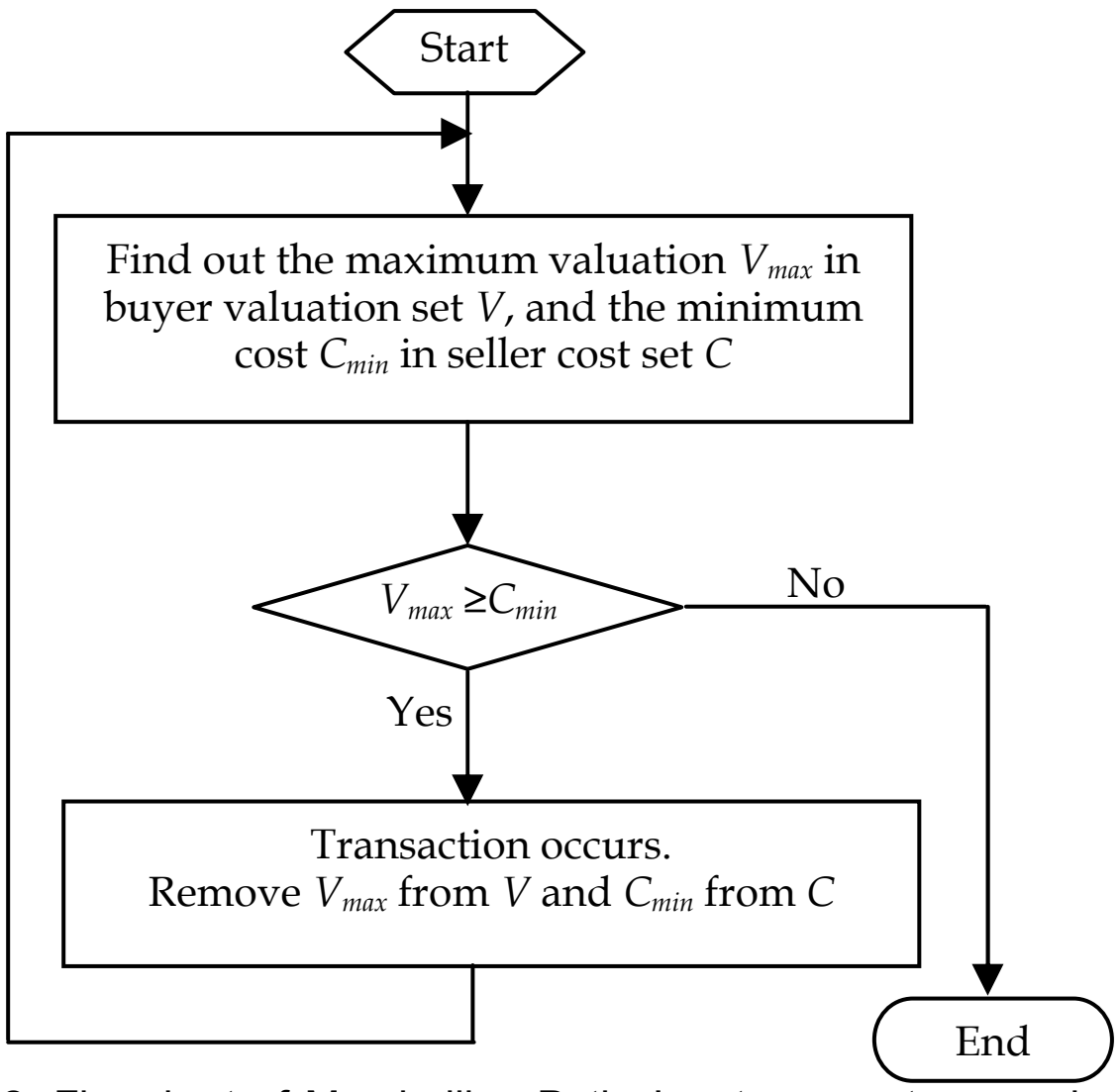

Fig. 2. Flowchart of Marshallian Path. Input parameters are buyers' values and sellers' costs. Output are successful trades.

According to the "invisible hand" theory in economics, after enough trading round buyers and sellers can achieve the equilibrium price and equilibrium quantity. The Marshallian path provides a theoretic description of how to achieve the equilibrium. Therefore the last transaction price in Marshallian path can be used as the approximation of the equilibrium price to evaluate performance of our cloud CDA market.

The Marshallian path is simply a sequence of trades from left to right along the supply and demand curves. If the maximum valuation of a buyer is equal to or more than the minimum cost of a seller, transaction occurs. The action is repeated until there is no valuation is equal to or more than a cost. The flowchart is shown in Fig. 2.

For example, the values of 16 buyers ranked in descending order are: 140, 125, 110, 95, 80, 85, 50, 45, 40, 35, 30, 25, 20, 15, 10, 5. And the costs of 16 sellers ranked in ascending order are: 30, 35, 40, 50, 55, 60, 65, 70, 75, 80, 85, 90, 95, 100, 105. Then the Marshallian path theory predicts the following sequence of trades as shown in table 1.

TABLE 1
TRADES PREDICTED BY MARSHALLIAN PATH

| Trade | 1 | 2 | 3 | 4 | 5 | 6 |
|---|---|---|---|---|---|---|
| Buyer | 140 | 125 | 110 | 95 | 80 | 65 |
| Seller | 30 | 35 | 40 | 45 | 50 | 55 |

No further trades are possible because the next buyer has a value (50) less than the seller's cost (60). Trade prices can vary anywhere between a buyer's unit value and the seller's unit cost. Thus, the initial possible range of prices is quite wide (30-140), but the possible range of prices is forced closer to the equilibrium. As trading progresses, the final trade (55-65) is constrained to be near the competitive equilibrium (55<P<60).

Thus an optimum allocation can be achieved using the Marshallian Path. To evaluate the efficiency of our cloud market and bidding strategy, it is a feasible way to use the allocation as a benchmark.

In the actual trading scenario, each trader can win one or more transactions, but in Marshallian Path each buyer or seller can only buy or sell once. We can compute the average surplus of one transaction. Therefore the market efficiency, $e_{market}$, can be rewritten as:

$$e_{market} = \left(\frac{\sum_i (S_{b,i}/n_b)}{\sum_i S_{b,i}^{MP}} + \frac{\sum_j (S_{s,j}/n_s)}{\sum_j S_{s,j}^{MP}}\right)/2 \qquad (5)$$

$S_{b,i}$ is total surpluses are won by buyer $i$, $n_b$ is maximum of trading rounds of all buyers in a CDA market. Accordingly $S_{s,j}$ and $n_s$ are values of sellers in a CDA market. $\sum_i S_{b,i}^{MP}$, $\sum_j S_{s,j}^{MP}$ are buyers' and sellers' total surpluses in the Marshallian Path respectively. Especially, $\sum_i (S_{b,i}/n_i)/\sum_i S_{b,i}^{MP}$ is buyers' efficiency, and $\sum_j (S_{s,j}/n_j)/\sum_j S_{s,j}^{MP}$ is sellers' efficiency. Our cloud computing market is a homogeneous market, so $e_s = e_{market}$.

Because the transaction price of the final trade in Marshallian Path, $p^{MP}$, is close to the equilibrium price $p^*$, the daily price volatility, $\alpha$, can be calculated with $p^{MP}$, as

$$\alpha = \frac{1}{p^{MP}} \sqrt{\frac{\sum_{i=1}^{N} (p_i - p^{MP})^2}{N}} \qquad (6)$$

In Section 5 (5) and (6) are used to evaluate our cloud CDA market and BH-strategy.

## 5 Belief-based Hybrid Strategy

Our e-CDA platform provides a feasible solution to cloud resource allocation and pricing. On such the platform, the selection of bidding strategies for the auction plays a very important role for each player to maximize its own profit. Furthermore, based on bidding strategies software agents can be designed to accomplish autonomous auctions for buyers and sellers.

Therefore we proposed a novel bidding strategy, Belief-based Hybrid Strategy (BH-strategy) appropriate for the cloud CDA mechanism. BH-strategy introduces an improved beliefs function and uses evolutionary programming to decide strategy dynamically. In this section, we describe BH-strategy in detail.

### 5.1 Beliefs Function

All Cloud users and CSPs attempt to maximize their surpluses in cloud computing market. However only when bids or asks are accepted and a transaction occurs, cloud users and CSPs can obtain surpluses. Therefore UBs or CSPs' agents must evaluate the probability of these bids or asks being accepted by other sellers or buyers, i.e. beliefs. It is a feasible method to form beliefs based on history trade records.

Gjerstad and Dickhuat [17] proposed a GD model, in which the trading activity is resulted from beliefs. Using the beliefs function, beliefs are formed on the basis of observed market data, including frequencies of asks, bids, accepted asks and accepted bids.

The work of Gjerstad and Dickhuat is flexible enough to respond quickly to changes in supply and demand conditions. But it is difficult for each buyer and seller to collect and calculate the history records (the recent $L$ trading rounds). Moreover, their simulations showed that for long memory lengths $(L \geq 8)$ the outcomes are similar to those with intermediate memory lengths $(4 \leq L \leq 7)$, but computation time increases significantly.

To reduce the computation time and costs, this paper introduced an improved beliefs function. It also forms beliefs based on history trading, but dose not need all details of the history bids and asks. It calculates the estimate of the competitive equilibrium price $p^*$ using recent transaction prices, and then forms the sellers' and buyers' beliefs according to the estimate of $p^*$.

Furthermore, our beliefs functions for both sellers and buyers are piecewise-defined. Generally transaction prices converge to the competitive equilibrium price $p^*$.

If at any time $o_{ask} \le p*$ the sellers should be less aggressive to submit a new ask $a$. Similarly, if at any time $o_{bid} \ge p*$ the buyers should be less aggressive to submit a new bid $b$. Therefore the beliefs function should be defined by two sub-functions, and each sub-function is applied to a certain interval of the main function's domain (a sub-domain). The main domain should be divided into two sub-domains at the point $p*$. As noted in section 3, the competitive equilibrium price $p*$ can not be known in advance, in our work the main domain is divided at the estimate of $p*$.

We can use the moving average method to calculate the estimate of $p*$ based on the history transaction prices. According to [11], the moving average is an objective analytical tool that gives the average value over a time frame spanning from the last transaction. It is sensitive to price changes over a short time frame, but over a longer time span, is less sensitive and filters out the high-frequency components of the signal within the frame. Different from [11], this paper uses the weighted moving average method to calculate the estimate of the competitive equilibrium price, denoted by $\hat{p}*$.

Therefore given a set of latest $HN$ transaction prices, (11) describes how to get $\hat{p}*$:

$$\hat{p}* = \frac{\sum_{i=T-HN+1}^{T} (w_i \times p_i)}{1+2+...+HN}, \qquad (7)$$
$$where\ w_{T-HN+1} = 1, ..., w_i = i-(T-HN), ..., w_T = HN$$

$(w_{T-HN+1}, ..., w_T)$ is the weight given to latest $HN$ transaction prices $(p_{T-HN+1}, ..., p_T)$, and $T$ is the latest transaction. If there is no transaction occurring before, then $\hat{p}* = 0$.

Our improved SELLERS' BELIEFS:

$$\hat{p}(a) = \begin{cases} 1 & if\ a \le \hat{p}* \\ \dfrac{TAG(a)+BG(a)}{TAG(a)+BG(a)+RAL(a)} & if\ Max \ge a > \hat{p}* \end{cases} \qquad (8)$$

Apparently the function $\hat{p}(a)$ should be monotonically non-increasing. In fact only when a seller submits an ask $a = Min$ ($Min$ is the **minimum bid** allowed in the market, defined in section 4.2), the probability of $a$ accepted by another buyer is 1, i.e. $\hat{p}(Min) = 1$. In GD model [17] the sellers' beliefs curve is smooth and decreasing at interval $(Min, Max]$, as shown in Fig. 4.

But in our work the bidding game is thought to consist of two stages: the aggressive stage and the unaggressive stage. Sellers take different actions in each stage. When $o_{ask} > \hat{p}*$ a seller can be more aggressive, i.e. in the aggressive stage the seller chooses the best ask based on belief. At interval $(\hat{p}*, Max]$ the seller's belief is computed using history bid and ask records. When $o_{ask} \le \hat{p}*$ a seller can be less aggressive, i.e. in the unaggressive stage the seller chooses the best ask without taking history records into account. At interval $[Min, \hat{p}*]$ the seller's belief is set to be a fixed value 1 to reduce computing costs. Therefore our sellers' beliefs function is a piecewise function, as shown in Fig. 4.

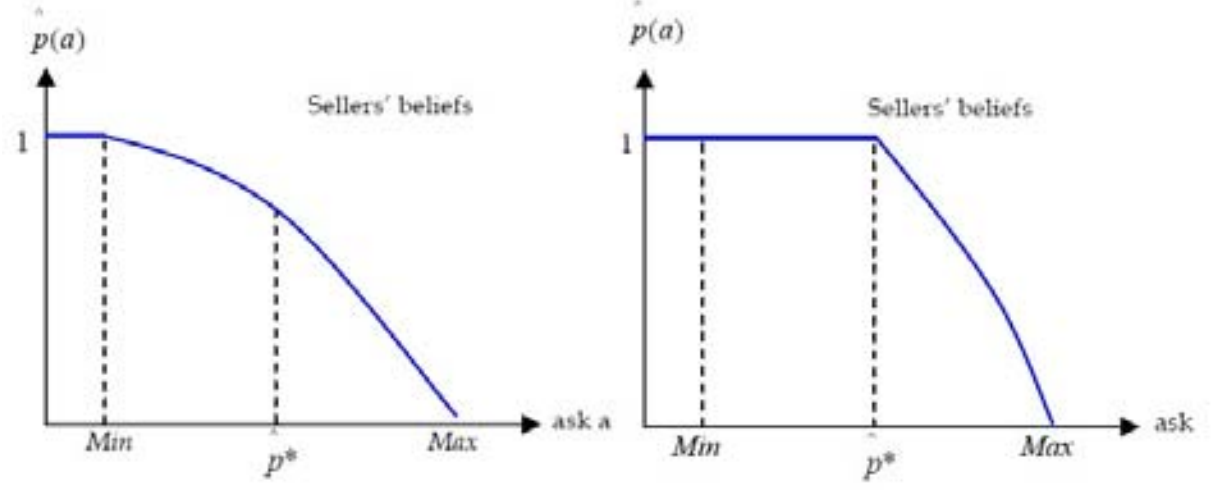

(a) GD strategy (b) BH strategy

Fig. 4. Sellers Beliefs Curves: (a) is sellers' beliefs curve of GD strategy, (b) is our improved sellers' beliefs curve for BH strategy.

To reduce computation complexity, the sellers' beliefs at interval $(\hat{p}*, Max]$ can be fitted a polynomial based on history transaction data. Our statistical experiments find that a cubic polynomial is good enough. Therefore our SELLERS' BELIEFS can be rewritten as:

$$\hat{p}(a) = \begin{cases} 1 & if\ a \le \hat{p}* \\ p_1 a^3 + p_2 a^2 + p_3 a + p_4 & if\ Max \ge a > \hat{p}* \end{cases} \qquad (9)$$

In (9) $p_1$, $p_2$, $p_3$, $p_4$ will be fixed by analyzing history transaction data in a given market populated by $N$ sellers and $M$ buyers.

Our improved BUYERS' BELIEFS:

$$\hat{q}(b) = \begin{cases} \dfrac{TBL(b)+AL(b)}{TBL(b)+AL(b)+RBG(b)} & if\ Min \le b < \hat{p}* \\ 1 & if\ b \ge \hat{p}* \end{cases} \qquad (10)$$

$\hat{q}(b)$ is monotonically non-decreasing. In fact only when a buyer submits a bid $b = Max$ ($Max$ is the **maximum ask** allowed in the market), the probability of $b$ accepted by another seller is 1, i.e. $\hat{q}(Max) = 1$. In GD model the buyers' beliefs curve is smooth and increasing at interval $(Min, Max]$, as shown in Fig. 5.

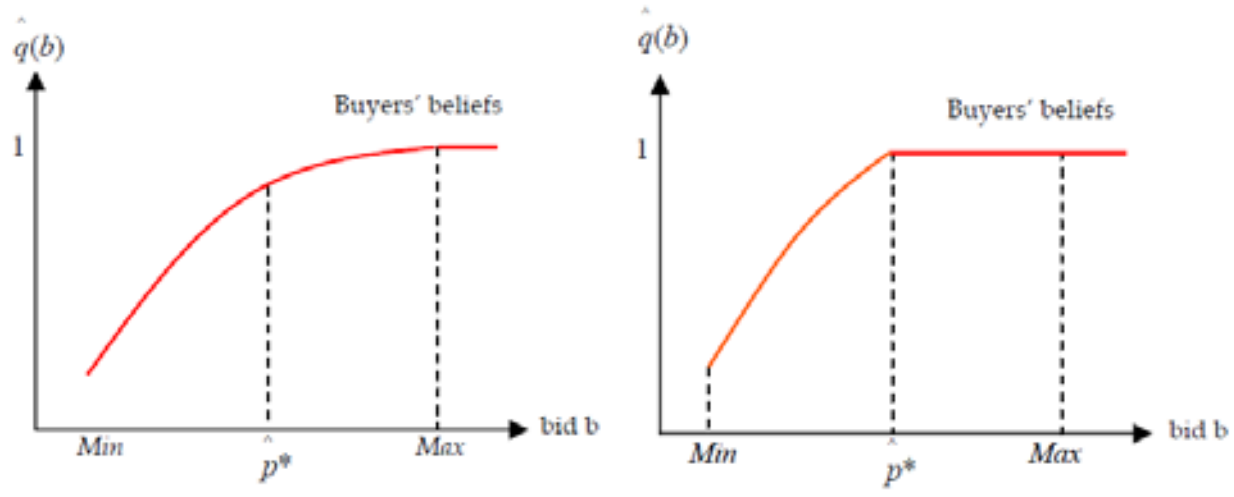

(a) GD strategy (b) BH strategy

Fig. 5. Buyers Beliefs Curves: (a) is buyers' beliefs curve of GD strategy, (b) is our improved buyers' beliefs curve for BH strategy.

In our two-stage bidding game, when $o_{bid} < \hat{p}*$ a buyer can be more aggressive, i.e. in the aggressive stage the buyer chooses the best ask based on belief. At interval $[Min, \hat{p}*)$ the buyer's belief is computed using history bid and ask records. When $o_{bid} \geq \hat{p}*$ a buyer can be less aggressive, i.e. in the unaggressive stage the buyer chooses the best ask without taking history records into account. At interval $[\hat{p}*, Max]$ the buyer's belief is set to be a fixed value 1 to reduce computing costs. Therefore our buyers' beliefs function is a piecewise function, as shown in Fig. 5.

Similarly, the buyers' beliefs at interval $[Min, \hat{p}*)$ can also be fitted a cubic polynomial as followed:

$$\hat{q}(b) = \begin{cases} q_1 b^3 + q_2 b^2 + q_3 b + q_4 & \text{if } Min \leq b < \hat{p}* \\ 1 & \text{if } b \geq \hat{p}* \end{cases} \quad (11)$$

## 5.2 BH-Strategy

Based on our improved beliefs functions, the buyers or sellers take different actions at different game stages.

### 5.2.1 Aggressive Stage

If $o_{ask} > \hat{p}*$, sellers should be in the aggressive stage. A seller should compute a best ask based on improved beliefs function. Sellers' Expected Surplus is defined as followed:

$$S_{s,j} = \max\{\max_a[(a - C_j)\hat{p}(a)], 0\} \quad (12)$$

For seller $j$, its best ask is $a_j$, maximizing $S_{s,j}$. Substituting (9) to (12), we have:

$$a_j = \arg\max\{(a - C_j)(p_1 a^3 + p_2 a^2 + p_3 a + p_4)\} \quad (13)$$

If $o_{bid} < \hat{p}*$, buyers should be in the aggressive stage. Buyers' Expected Surplus is defined as the following:

$$S_{b,i} = \max\{\max_b[(V_i - b)\hat{q}(b)], 0\} \quad (14)$$

For buy $i$, its best bid is $b_i$ maximizing $S_{b,i}$, i.e.

$$b_i = \arg\max\{(V_i - b)(q_1 b^3 + q_2 b^2 + q_3 b + q_4)\} \quad (15)$$

### 5.2.2 Unaggressive Stage

If $o_{ask} \leq \hat{p}*$, sellers should be in the unaggressive stage. Similarly, if $o_{bid} \geq \hat{p}*$, buyers should be in the unaggressive stage. When in this stage a seller or buyer submits a new ask or bid, which means it will accept a worse price than history traders did.

For example a seller $j$ submits an ask $a < o_{ask}$. If the ask $a$ is accepted, the seller $j$ can be thought to lose the surplus $Loss_j$ compared with history trading price:

$$Loss_j = (\hat{p}* - C_j) - (a - C_j) = \hat{p}* - a \quad (16)$$

We assume it is rational for a seller to decide the best ask $a$ based on whether it has been submitted or accepted in history trading round. So does a buyer.

Furthermore, in our cloud CDA mechanism asks or bids submitted by traders must be subject to constraints (2) and (3). When $o_{ask} \leq \hat{p}*$, the interval $[C_j, o_{ask})$ has already been too small to provide the seller j with more choice. Similarly when $o_{bid} \geq \hat{p}*$, the interval $(o_{bid}, V_i)$ has been also too small.

Therefore in unaggressive stage a seller just submits a random ask as followed:

$$a_j \sim U(C_j, o_{ask}) \quad (17)$$

$U(C_j, o_{ask})$ is the uniform distribution.

A buyer submits a random bid as followed:

$$b_i \sim U(o_{bid}, V_i) \quad (18)$$

Similarly, $U(o_{bid}, V_i)$ is the uniform distribution.

### 5.2.3 First Round

In our two-stage bidding game, the sellers take aggressive or unaggressive actions depending on whether the current $o_{ask}$ is greater than $\hat{p}*$ or not. So do the buyers.

But in the first trade round no transaction occurs, so $\hat{p}*$ can not be estimated. The seller or buyer has no history trading information other than its limit price. If the seller $j$ submits a too low ask, it can transact at a not very profitable price (with respect to $p*$). Therefore, it starts with high asks that progressively approach the maximum of its cost price $C_j$ and the outstanding bid $o_{bid}$ to explore the market. Similarly the buyer $i$ submits a bid towards the minimum of its valuation $V_i$ and the outstanding ask $o_{ask}$ to explore the market.

Therefore we defined bid rules in the first trading round by adopting method of [11]. The seller $j$ should submit $a_j$ in the first trading round as following:

$$a_j = o_{ask} - \frac{o_{ask} - \max\{C_j, o_{bid}\}}{\eta} \quad (19)$$

And buyer $i$ should submit $b_i$ in the first trading round as following:

$$b_i = o_{bid} + \frac{\min\{V_i, o_{ask}\} - o_{bid}}{\eta} \quad (20)$$

Thus the bid-ask spread so is reduced with an exponentially decreasing trend determined by $\eta$ and its limit price. Here a low $\eta$ implies a faster rate of convergence of bids or asks until they are matched at a transaction price. Conversely, a high $\eta$ implies a more conservative bidding approach and a slower convergence. According to [11], $\eta = 3$ was observed to be a good compromise. Therefore in the first trading round we set $\eta = 3$ and compute ask and bid using (19) and (20).

#### *5.2.4 BH-strategy*

Therefore the Belief-based Hybrid Strategy can be described as the following:

**Bidding Strategy for Seller** j:
**If** ( $C_j \geq o_{ask}$ ) submit no ask
**Else**
**If** (first trading round, $\hat{p}* = 0$ ) submit an ask given by (19)
**Else**
**If** ( $o_{ask} > \hat{p}*$ ) /*aggressive stage*/
submit an ask computed by (13)
**Else** /*unaggressive stage*/
submit an ask given by (17)
**End If**
**End If**
**End If**

**Bidding Strategy for Buyer** i:
**If** ( $V_i \leq o_{bid}$ ) submit no bid
**Else**
**If** (first trading round, $\hat{p}* = 0$ ) submit a bid given by (20
**Else**
**If** ( $o_{bid} < \hat{p}*$ ) /*aggressive stage*/
submit a bid computed by (15)
**Else** /*unaggressive stage*/
submit a bid given by (18)
**End If**
**End If**
**End If**

As shown above, at the first trading round the BH buyers and sellers have no information of history trading, so they submit orders based on the current outstanding orders and their limit prices.

From the second trading round $\hat{p}*$ can be computed and BH buyers and sellers take different bidding strategies according to $\hat{p}*$.

## 6 Simulation and Evaluation

In this section, we first detail the design of simulation for analyzing the strategic interaction of the BH-strategy in cloud CDA markets. Especially we compare our strategy with ZI strategy, GD strategy and AA strategy. Then the actual empirical study of performance is given.

### 6.1 Simulation Design

We simulate scenarios with three different kinds of scales to evaluate our strategy in Matlab.

In the small simulation scenario the market is populated with a set of 10 buyers and 10 sellers on a same scale with [11]. In the large simulation scenario, there are 100 buyers and 100 sellers in the market. In actual cloud computing markets there are often more cloud user brokers than CSPs' agents. Therefore we design the third scenario, the asymmetric scenario, populated with a set of 1000 buyers and 10 sellers.

For each scenario we simulate 30 trading days, and the length of each trading day is 1000. After 1000 steps (a bid or an ask is submitted), the trading is over and all bids and asks are cleared.

The buyers' values and sellers' costs are uniformed random numbers between 1 and the amount of sellers. In the small simulation scenario $V, C \sim U(1,10)$. In the large simulation scenario $V, C \sim U(1,100)$. In the asymmetric scenario $V, C \sim U(1,10)$.

The minimum increment of bid or ask $\Delta$ is an infinitely small quantity, i.e. a buyer/seller can submit a new bid/ask greater than $o_{bid}$ /$o_{ask}$. The minimum bid allowed in the market *Min=0*. The maximum ask allowed in the market $Max = 10 \times N_s$ ($N_s$ is amount of sellers). Table 2 provides simulation parameters in the three scenarios.

TABLE 2
PARAMETERS IN THREE SIMULATION SCENARIOS

| Parameter | Small | Large | Asymmetric |
|---|---|---|---|
| Buyers | 10 | 100 | 1000 |
| Sellers | 10 | 100 | 10 |
| V | U(1,10) | U(1,100) | U(1,10) |
| C | U(1,10) | U(1,100) | U(1,10) |
| Δ | ε | ε | ε |
| Min | 0 | 0 | 0 |
| Max | 100 | 1000 | 100 |
| $L_{TD}$ | 1000 | 10000 | 10000 |
| Trading Day | 30 | 30 | 30 |

*$\varepsilon$ is an infinitely small quantity. $L_{TD}$ is step amount in a trading day.*

The above three scenarios include simulations used frequently in related works, such as [11] used the first scenario, and [20] also applied uniformed random numbers to simulate values and costs. Furthermore, they represent the feature of real cloud markets, which will be addressed in section 6.4.

### 6.2 Benchmark

To evaluate BH-strategy, we will compare it with ZI strategy, GD strategy and AA strategy.

ZI strategy by Gode and Sunder [15] makes a uniformed, but profitable decision that is not based on observed market information. A ZI buyer submits an offer drawn from a uniform distribution between the minimum bid allowed in the market and its valuation, and a ZI seller between its cost and the maximum ask allowed in the market. Although ZI is a nonintelligent strategy, it is sufficient to raise the allocative efficiency of CDA more than 90%. Therefore ZI model is usually used as the benchmark when evaluating the CDA biding strategies [11, 16, 18, 20].

GD strategy [17] is an expected profit-maximizing and belief-based strategy. It calculates its belief that a bid or ask will be accepted in the market based on a set of the latest transactions and the expected profit associated with such bids and asks. Then, the bid or ask that maximizes the expected profit is submitted in the market. Our BH-strategy improves the beliefs function of GD, so we also compare our work with GD.

AA (Adaptive Aggressive) strategy [11] is based on both short and long-term learning that allow agents to modify their bidding behavior to be efficient in a market. For the short-term learning, the agent updates the aggressiveness of its bidding behavior based on market information observed after any bid or ask appears in the markets. The long-term learning determines how this aggressiveness factor influences an agent's choice in the market, and is based on market information observed after every successful transaction. As shown in [11] AA has better performance than other strategies, so we also use AA as the benchmark. In our simulation, sellers' costs and buyers' valuations are generated every trading day, so we only compare short-term learning AA strategy with our BH-strategy.

### 6.3 Evaluation

Successful transactions, surpluses, market efficiency (i.e. strategy efficiency in homogeneous markets), daily price volatility and executing time are evaluated to assess the performance of our cloud CDA market and BH-strategy. These are also the general evaluation criteria in related researches on auction mechanisms [11]. As stated above, we design three simulation scenarios: small, large and asymmetric scenario. We will measure these five criteria in each scenario respectively.

#### *6.3.1 Successful Transactions*

The number of successful transactions in one trading day is basic measurements of the auction efficiency. Because our cloud CDA market is a model with constraints, if transaction occurs both the winning seller and buyer obtain surpluses. It means that more successful transactions cause more total surpluses.

In the small and the large scenario ZI ranks first and BH second, and in the asymmetric scenario BH ranks first (see Appendix A). To illustrate them clearly, table 3 gives the average of daily successful transactions in these three scenarios.

TABLE 3
THE AVERAGE OF DAILY TRANSACTIONS

| Strategy | Small | Large | Asymmetric |
|---|---|---|---|
| ZI | 452 | 8326 | 6606 |
| GD | 228 | 677 | 3535 |
| AA | 305 | 4570 | 6560 |
| BH | 367 | 6679 | 7685 |

We observe that BH has overwhelmingly more successful transactions than GD, and also a little more than AA in all the three scenarios. ZI is a non-intelligent bidding strategy, so it has usually more successful transactions. However, in the asymmetric scenario BH has more transactions than ZI.

#### *6.3.2 Surpluses of Sellers and Buyers*

Every market mechanism aims at maximizing surpluses or profits attained by sellers and buyers in the market. We select three kinds of surplus criteria: daily sellers' surplus, buyers' surplus and total surpluses (the sum of sellers' surplus and buyers' surplus). Daily surpluses depend on the number of transactions and prices. Therefore it takes a global view of 4 bidding strategies by empirical study on these surplus criteria.

Fig. 6 offers total surpluses (sellers' surplus plus buyers' surplus) of ZI, GD, AA and BH in three scenarios. The details of the sellers' and the buyers' surplus are given in Appendix A respectively.

According to the experiment results, AA strategy obtains the most sellers' surplus of all and BH strategy obtains the most buyers' surplus. But considering sellers' and buyers' surpluses together, BH obtains the most in all the three scenarios.

#### *6.3.3 Efficiency*

As defined in Section 3, market efficiency is the ratio of the surpluses of all traders to the possible maximum surpluses that would be obtained in a centralized and optimum allocation.

Table 4 gives the average market efficiency:

TABLE 4
THE AVERAGE MARKET EFFICIENCY

| Strategy | Small | Large | Asymmetric |
|---|---|---|---|
| ZI | 0.8500 | 0.1430 | 0.7215 |
| GD | 0.8636 | 0.3081 | 0.6932 |
| AA | 0.9137 | 0.3420 | 0.8239 |
| BH | 0.9230 | 0.3542 | 0.8305 |

As shown in table 4, the average market efficiency of BH ranks first in all three scenarios. The details of daily market efficiency are shown in appendix A.

#### *6.3.4 Daily Price Volatility*

Daily price volatility, $\alpha$ , shows how the transaction prices converge to the equilibrium price. Fig. 16 gives the daily price volatility of ZI, GD, AA and BH in the three scenarios.

As shown in Fig. 7 average $\alpha$ of our BH is the smallest in the small scenario, and average $\alpha$ of AA is the smallest in the other two scenarios. Although AA is doing best on average daily price volatility, our BH is doing much better than GD and ZI.

To demonstrate how transaction prices distribute, Fig. 13 gives transaction prices in one trading day. In this figure *MP* denotes equilibrium price found by Marshallian Path. Therefore all dots of *MP* form a line, called *MP-line*. The faster transaction prices converge to *MP-line*, the smaller the daily price volatility is.

Fig. 8 shows how BH converges to equilibrium price. Although the daily price volatility of BH is not obviously smaller than AA, the difference among transaction prices in one trading day of BH is smaller than the others.

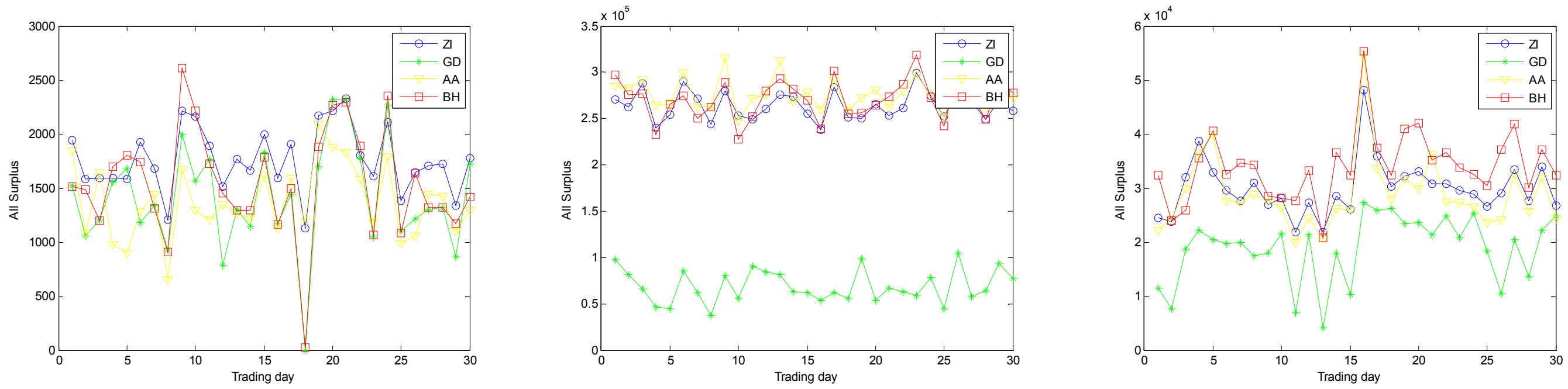

(a) Small Simulation Scenario (b) Large Simulation Scenario (c) Asymmetric Simulation Scenario

Fig. 6. Total Surpluses Evaluation

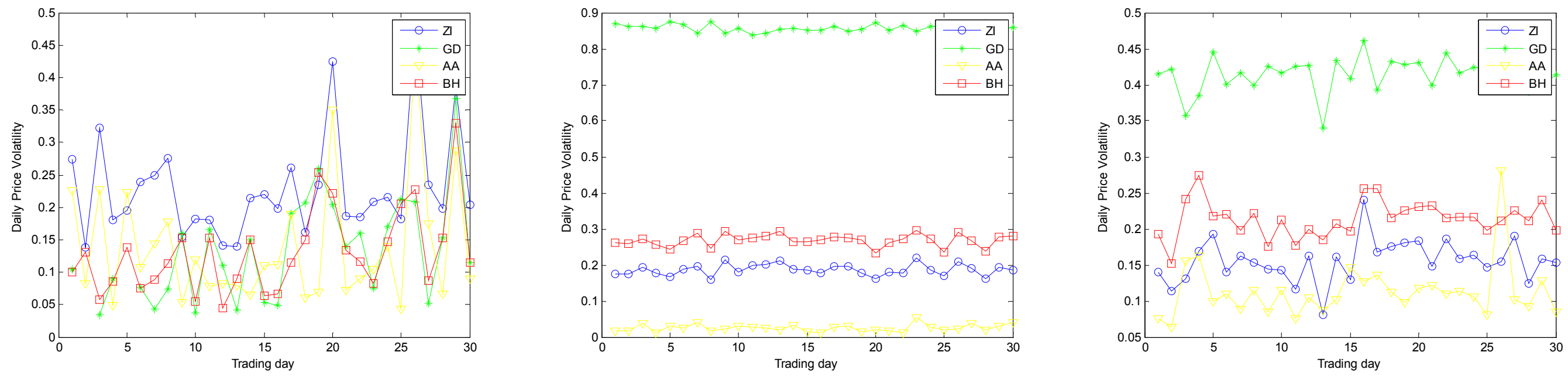

(a) Small Simulation Scenario (b) Large Simulation Scenario (c) Asymmetric Simulation Scenario

Fig. 7. Daily Price Volatility

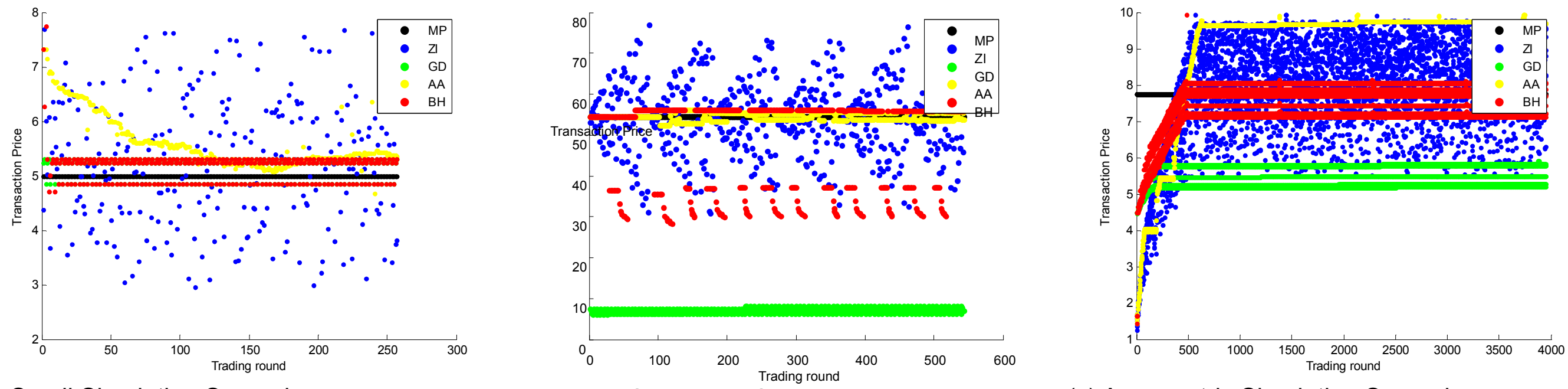

(a) Small Simulation Scenario (b) Large Simulation Scenario (c) Asymmetric Simulation Scenario

Fig. 8. Transaction Prices in One Trading Day

### *6.3.5 Executing Time*

When a trader adopts a bidding strategy, it will execute the strategy algorithm to find the best offer. Therefore the algorithm time complexity is a key factor of the trading costs. Especially on real cloud e-CDA platform, users and CSPs may adopt software agents to trade automatically, so a low time cost algorithm is very important.

We use simulation executing time in Matlab to evaluate it. The start and the end time are serial date numbers defined in Matlab, and their difference is the executing time. Fig. 9 provides the executing time of ZI, GD, AA and BH in the three simulation scenarios.

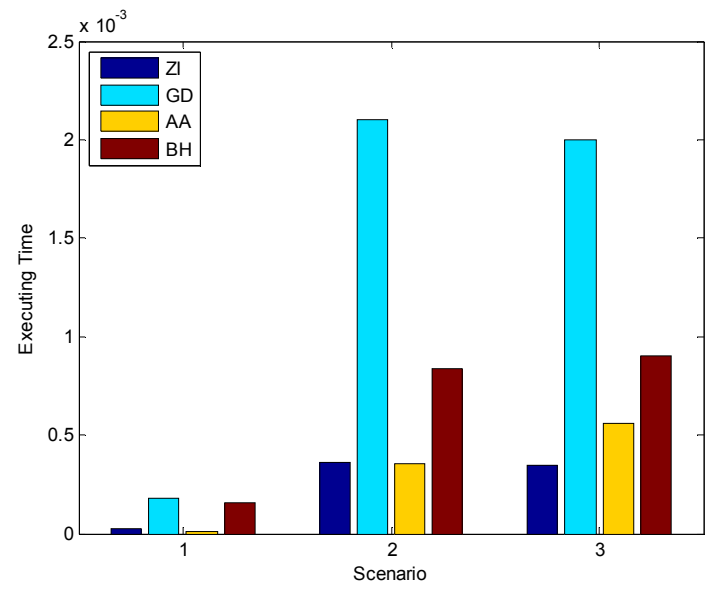

Fig. 9. Executing Time: x=1, 2, 3 represent small scenario, large scenario and asymmetric scenario.

As shown in Fig. 9, the executing time of GD is the longest in all scenarios because of finding a local maximum of expected surplus function. Our BH reduces the executing time comparing to GD. Although BH needs a longer executing time than AA, BH has higher efficiency than AA. In experiments it takes some time for MATLAB to generate random bids, so the executing time of ZI is not short. But in actual auction, if a trader takes ZI strategy it does not need to compute.

### *6.3.6 Summary of Simulations*

The summary of above the simulation is listed in table 5:

TABLE 5
THE SUMMARY OF SIMULATIONS

| Criteria | Experiment Results |
|---|---|
| Successful Transactions | BH has more successful transactions than GD and AA in the three scenarios. Although in the small and the large scenario BH is less good than ZI, in the asymmetric scenario BH has more transactions than ZI. |
| Surpluses | AA strategy obtains the most sellers' surplus of all and BH strategy obtains the most buyers' surplus. But considering sellers' and buyers' surpluses together, BH obtains the most in three scenarios. |
| Efficiency | The average market efficiency of BH is the largest in the three scenarios |
| Daily Price Volatility $\alpha$ | Average $\alpha$ of our BH is the smallest in the small scenario, and AA's is the smallest in the other two scenarios. Although AA is doing best on average daily price volatility, our BH is doing much better than GD and ZI. |
| Daily Price Difference | Although average daily price volatility of BH is not obviously smaller than AA, the difference among transaction prices in one trading day of BH is smaller than the others. |

By analyzing the results we can conclude that BH performs the best in all the scenarios on surpluses, market efficiency and daily price difference. For successful transactions, BH ranks the second while ZI the first. But ZI is a nonintelligent strategy and BH is overwhelmingly better than ZI in other aspects. For daily price volatility, BH ranks the second while AA the first. But it only shows the difference between transaction prices and equilibrium prices estimated by Marshallian Path. Because the difference among transaction prices in one trading day of BH is smaller than the other strategies, it is also acceptable.

## 6.4 Feasibility in Real Cloud Markets

Our simulations only consist of 3 scenarios, but the results also prove the feasibility of our cloud CDA mechanism in real cloud markets. Though there are more users in real cloud markets than our simulations, in each auction not all users participate in. Therefore, the scale of each auction in real cloud markets is close to one of our simulation scenarios. For example, when a CSP published one of its idle VM instances, maybe only several users took part in the auction to bid it.

To illustrate why the cloud CDA mechanism is feasible, we analyze features of commodities and buyers in real cloud markets as the following:

**1. Homogeneous Commodities:** Sellers in cloud markets are CSPs, who provide users with computing and storage resources, i.e. commodities. On our e-CDA platform auction commodities should be traded with uniform units, i.e. a standard measurement approach to computing and storage. Despite the difference of the physical machine and storage medium among CSPs, a system of unit for cloud resource can be predefined and approved by both buyers and sellers.

The unit for computing service can be a VM instance with fixed CPU, I/O, memory, and network bandwidth. For instance, Amazon EC2 defined "EC2 Compute Units" or ECUs, as a measure of virtual computing power. The e-CDA platform can also define one VM instance as a standard unit of computing power, named Compute Unit (CU). Therefore a computing commodity can be described as *40 CUs*. Owing to the development of Virtualization Technology, it is feasible for CSPs to create VM instances and operating costs of the VM management will be considerably reduced.

The unit for storage service is also designed in the same way. Like Amazon S3, the Storage Unit (SU) can be 1TB/month with fixed number of GET, PUT, COPY, POST, or LIST requests. For example a storage commodity can be represented as *10 SUs*. Similarly, current storage technology makes it easy to allocate a maximum amount of storage space and requests that a user can use.

With the above standard system of units for computing and storage, the commodities traded on the e-CDA platform can be deemed as homogeneous. Therefore the cloud CDA mechanism can be applied.

**2. Heterogeneous User Demands:** Buyers in cloud markets are cloud users, who may differ in their service requirements and valuation parameters. According to [9, 22], cloud users can be classified into 2 categories: Job-oriented users and Resource-aggressive users.

Job-oriented users often have batch jobs, which need a certain computation to be carried out, and the computation time scales with the allocated resources. Such jobs are common, for example, in scientific and business computing. Such a user can submit his/her demand of a computing job, including size, available time and deadline, on the e-CDA platform. Then the platform initiates an auction for it. The user can submit bids less than the budget of the job. If the CSPs have idle resources to satisfy the demand, they can submit asks more than their costs. Other users also can submit bids for it.

Resource-aggressive users often have fixed duration applications. In certain application classes, cloud resources may be secured for fixed periods of time. For example, a small ICP (Internet Content Provider) needs a VM instance to host a web site for one year, or a user buys 1TB storage for six months as online backup. Such demands can also be described as concise request to fit for publishing on the e-CDA platform.

In summary the features of commodities and users in real cloud markets make it feasible adopting CDA to allocate cloud resources. Cloud resources can be measured with uniform units and heterogeneous user demands can be described as concise request, therefore the cloud users and CSPs can trade on the e-CDA platform.

## 7 Conclusion

Cloud computing is a new and promising paradigm delivering IT services as computing utilities. In such enormous computing resources markets, CDA is an effective method to decentralized allocation.

In this paper a cloud Continuous Double Auction (CDA) mechanism is brought for cloud resources allocation. We define the market rules to match orders and facilitate trading, and introduce evaluation criteria to measure performance. Furthermore, we design an electronic bidding platform to implement this mechanism in the cloud environment. Then we develop a novel bidding strategy for cloud CDA, BH-strategy, which is a two-stage game bidding strategy based on improved beliefs function. In the aggressive stage it decides the trading activity based on the improved beliefs function. In the unaggressive stage it chooses the best action according to static rules.

By simulation we compare our BH-strategy with other typical strategies in successful transactions, surpluses, market efficiency, daily price volatility and executing time in three simulation scenarios: small, large and asymmetric scenario. BH-strategy has better performance on surpluses, successful transactions and market efficiency. Especially, it largely improves market efficiency in the asymmetric scenario which is closer to the actual cloud market.

Furthermore, we discuss the feasibility of our CDA cloud mechanism in real cloud markets. The analysis demonstrates it is a feasible market for cloud computing resource allocation.

## Appendix A. Empirical Study in the Three Scenarios

In section 6, we benchmarked our BH-strategy against the ZI, GD and AA strategies for the three scenarios. Here, we provide the details of daily successful transactions, sellers' surplus, buyers' surplus and market efficiency which have not been presented in the main body of the paper.

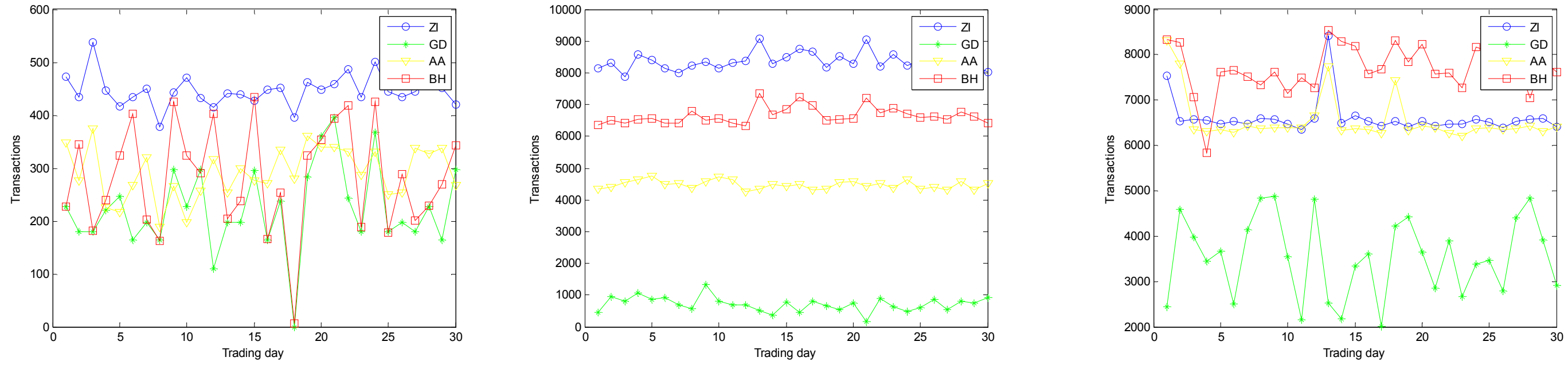

(a) Small Simulation Scenario (b) Large Simulation Scenario (c) Asymmetric Simulation Scenario

Fig. A.1. Successful Transactions

As shown above in the small and the large scenario ZI ranks first and BH second, and in the asymmetric scenario BH ranks first.

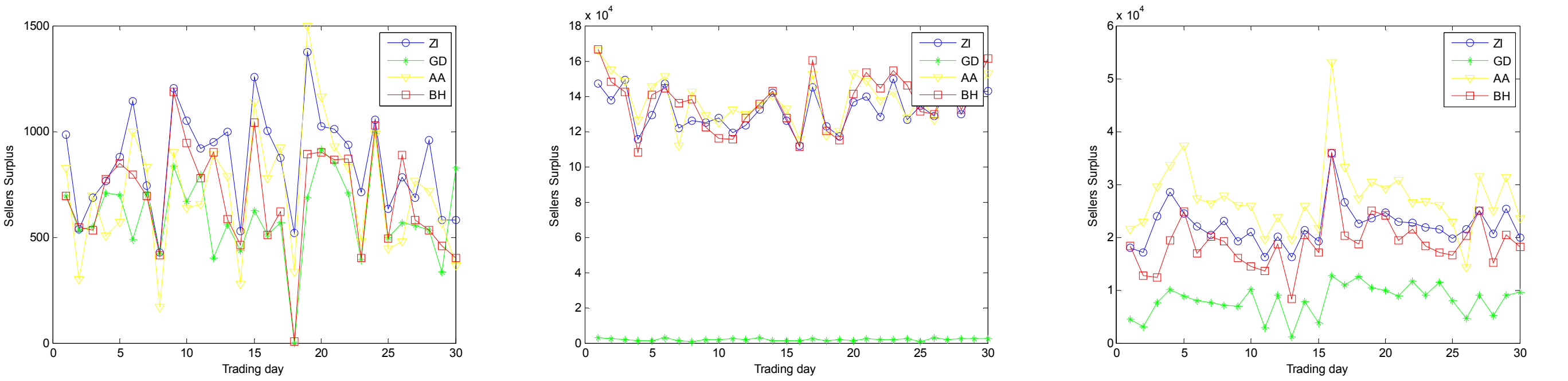

(a) Small Simulation Scenario (b) Large Simulation Scenario (c) Asymmetric Simulation Scenario

Fig. A.2. Sellers Surplus Evaluation

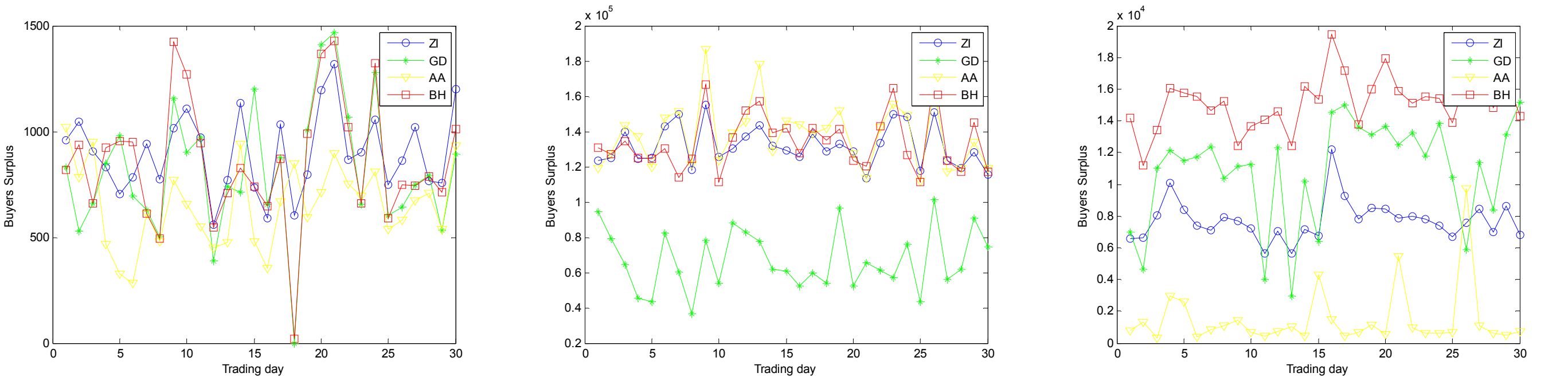

(a) Small Simulation Scenario (b) Large Simulation Scenario (c) Asymmetric Simulation Scenario

Fig. A.3. Buyers Surplus Evaluation

Fig. A.2 gives the sellers' surpluses of ZI, GD, AA and BH in these three scenarios. Fig. A.3 shows the buyers' surpluses of ZI, GD, AA and BH in these three scenarios. According to the experiment results, AA strategy obtains the most sellers' surpluses of all and BH strategy obtains the most buyers' surpluses. But considering sellers' and buyers' surpluses together, BH obtains the most in all the three scenarios.

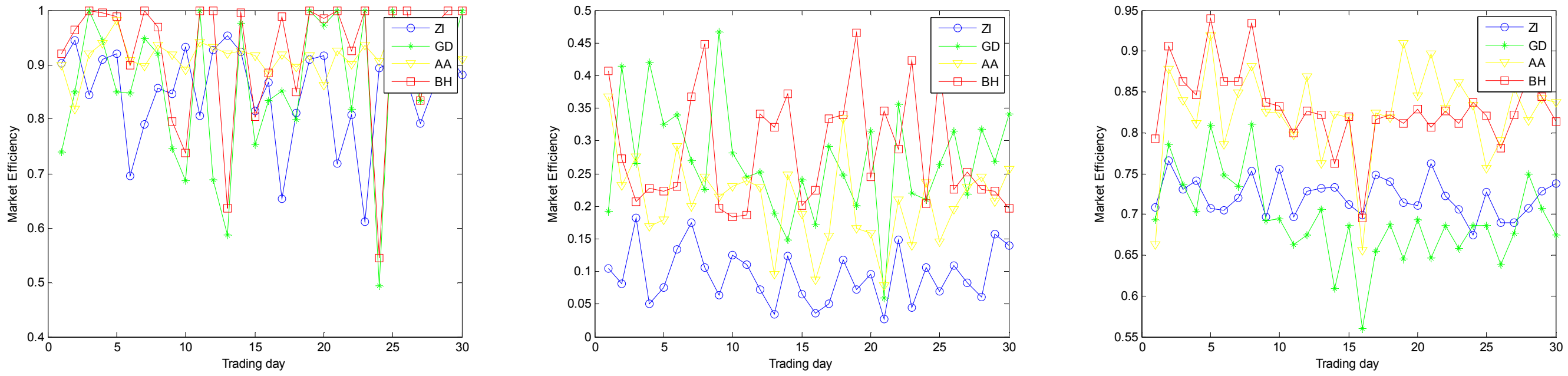

(a) Small Simulation Scenario (b) Large Simulation Scenario (c) Asymmetric Simulation Scenario

Fig. A.4. Market Efficiency Evaluation

Fig. A.4 provides the daily efficiency of ZI, GD, AA and BH in all the three scenarios. The average market efficiency of BH ranks first in all three scenarios.

## Acknowledgment

The authors wish to thank the reviewers for their valuable comments that will enable us to improve the paper. This work was supported in part by China National Science and Technology Ministry (2011BAK08B05-02), China National Science & Technology Major Project (2012ZX03005001), National Natural Science Foundation of China (61170292, 60970104) and China 973 project (2012CB315803).

**Xuelin Shi** received her PhD in CS from Beijing Institute of Technology in 2005, and received her MS and BS in CS from Beijing University of Chemical Technology in 2002 and 1999 respectively. From 2005 to 2006 she was a software engineer of Beijing Founder Company. From 2006 to 2012 she was an assistant professor of Beijing University of Chemical Technology. Since 2012 she has been an assistant researcher in the Department of Computer Science and Technology at the Tsinghua University. Her current research interests include cloud computing, game theory, network utility optimization.

**Ke Xu** (M'02-SM'09) received the BS, MS, and PhD degrees in computer science from Tsinghua University, China, in 1996, 1998, and 2001, respectively. Currently he is a full professor in the Department of Computer Science of Tsinghua University. His research interests include next generation Internet, traffic management, switch and router architecture, and P2P and overlay network. He is a senior member of the IEEE and a member of the ACM.

**Jiangchuan Liu** (S'01-M'03-SM'08) received the BEng degree from Tsinghua University, Beijing, China, in 1999, and the PhD degree from The Hong Kong University of Science and Technology in 2003, both in computer science. He is a recipient of Microsoft Research Fellowship (2000), Hong Kong Young Scientist Award (2003), and Canada NSERC DAS Award (2009). He is a co-recipient of the Best Student Paper Award of IWQoS'2008, the Best Paper Award (2009) of IEEE ComSoc Multimedia Communications Technical Committee, and Canada BCNet Broadband Challenge Winner Award 2009. He is currently an Associate Professor in the School of Computing Science, Simon Fraser University, British Columbia, Canada, and was an Assistant Professor in the Department of Computer Science and Engineering at The Chinese University of Hong Kong from 2003 to 2004. His research interests include multimedia systems and networks, wireless ad hoc and sensor networks, and peer-to-peer and overlay networks. He is a Senior Member of IEEE and a member of Sigma Xi. He is an Associate Editor of IEEE Transactions on Multimedia, and an editor of IEEE Communications Surveys and Tutorials. He is TPC Vice Chair for Information Systems of IEEE INFOCOM'2011.

**Yong Wang** received the PhD degree in industrial economics from Peaking University in 2003, and received the MS degree in western economics from Nankai University in 1999, and received the BS degree in political economics from Nankai University in 1996. Currently he is an associate professor in the School of Social Sciences of Tsinghua University. His research interests include Microeconomics, industrial organization, theory of the firm, and corporate finance.